\documentclass[preprint]{elsarticle}

\usepackage{amssymb, latexsym, amsthm, amsmath, epsfig, mathtools, todonotes, booktabs, cite, graphicx}
\usepackage[singlelinecheck=false]{caption}

\journal{Theoretical Population Biology}

\newcommand\E{\operatorname{E}}

\newcommand\given{{\,|\,}}

\newcommand\eg{{\it e.g.,}}

\newcommand\ie{{\it i.e.,}}

\newcommand\etc{{\it etc}}

\begin{document}

\begin{frontmatter}

\title{Maximum likelihood (ML) estimators for scaled mutation parameters with a strand symmetric mutation model in equilibrium}
\author[address1]{Claus Vogl\corref{correspondingauthor}}
\cortext[correspondingauthor]{Corresponding author}
\ead{claus.vogl@vetmeduni.ac.at}
\author[address2]{Lynette Caitlin Mikula}
\ead{lcm29@st-andrews.ac.uk}

\address[address1]{Department of Biomedical Sciences, Vetmeduni Vienna, Veterin\"arplatz 1, A-1210 Wien, Austria}
\address[address2]{Centre for Biological Diversity, School of Biology, University of St.\ Andrews, St Andrews KY16 9TH, UK}

\begin{abstract}
With the multiallelic parent-independent mutation-drift model, the equilibrium proportions of alleles are known to be Dirichlet distributed. A special case is the biallelic model, in which the proportions are beta distributed. A sample taken from these models is then Dirichlet-multinomially or beta-binomially distributed, respectively. Maximum likelihood (ML) estimators for the mutation parameters of the biallelic parent-independent mutation model are available via an expectation maximization algorithm. Assuming small scaled mutation rates, the distribution of a sample of size $M$ can be expanded in a Taylor series of first order. Then the ML estimators for the two parameters in the biallelic model can be expressed using the site frequency spectrum. In this article, we go beyond parent-independent mutation and analyse a strand-symmetric mutation model with six scaled mutation parameters that deviates from parent independent mutation and, generally, from detailed balance. We derive ML estimators for these six parameters assuming mutation-drift equilibrium and small scaled mutation rates. This is the first time that ML estimators are provided for a mutation model more complex than parent-independent mutation. 
\end{abstract}


\end{frontmatter}


\section{Introduction}

With the parent-independent mutation-drift model, theoretical results for the estimation of mutation parameters are available: The proportions of alleles of a biallelic model in equilibrium are beta distributed \citep{Wright31}. Data for the inference of parameters are usually in the form of site frequency spectrum (also called allele frequency spectrum) data taken from one population. Such data are a vector of allele frequencies for samples of $M$ haploid individuals at a total of $L$ sites (or loci) whereby there are  $L_0,L_1,\dots,L_M$ sites with $y=0,1,\dots, M$ alleles of the focal type. Given a sample of size $M$ and a binomial sampling distribution conditional on the allele proportion, one obtains a beta-binomial distribution for the allele frequencies. Maximum likelihood (ML) estimators can be constructed from the beta-binomial distribution via an expectation maximization algorithm, where a root of a polynomial of the order of the sample size needs to be evaluated at each iteration \citep{Vogl14}. With a multiallelic parent-independent mutation model, a Dirichlet-multinomial distribution for the allele frequencies follows analogously.  

In the limit of small scaled mutation rates, it is convenient to reparametrize the two parameters of the beta distribution for equilibrium allele proportions with the mutation bias $\alpha$ and the overall mutation rate $\theta$. Expanding the beta-binomial distribution in a Taylor series up to first order of the scaled mutation rate $\theta$ then results in a relatively simple equilibrium sample distribution \citep{Vogl14}:
\begin{equation}\label{eq:Vogl_Bergman}
 {\Pr}(y\given M,\alpha,\theta\to 0)=\begin{cases}
 (1-\alpha)\big(1-\alpha\theta H_{M-1}\big) + O(\theta^2)&\text{for $y=0$,}\\
 \alpha(1-\alpha)\theta\,\frac{M}{y(M-y)} + O(\theta^2) &\text{for $1\leq y\leq (M-1)$,}\\
 \alpha\big(1-(1-\alpha)\theta H_{M-1}\big) + O(\theta^2) &\text{for $y=M$}\,.
 \end{cases}
\end{equation}
whereby $H_m$ denotes the $m$th harmonic number. Note that since $H_{M-1}$ approaches infinity logarithmically, the sample size must stay below a limit of approximately $M<e^{\frac{1}{\max(1-\alpha,\alpha)\theta}}$. Furthermore, simulations have shown that the mutation rates for the first order Taylor series expansion must be $2\alpha(1-\alpha)\theta\leq 0.025$ \citep{VoglClemente12}; this upper boundary for the scaled mutation rate was further substantiated by \citet{SchrempfHobolth17}. It is higher than the scaled mutation rate of most eukaryotes \citep{Lynch16}. With this approximation, simple ML estimators given site-frequency spectrum data are obtained: 
\begin{equation}
    \hat\alpha=\frac{L_M+L_p}{L}
\end{equation}
and
\begin{equation}
    \hat\theta=\frac{1}{2\alpha(1-\alpha)}\frac{L_p}{H_{M-1}L}\,,
\end{equation}
respectively \citep{Vogl14}. 

\citet{VoglClemente12} proposed a Moran model, in which mutations are restricted to monomorphic states. In this so-called boundary-mutation Moran model, the same equilibrium sample distribution is obtained as with the first order Taylor series expansion in formula~(\ref{eq:Vogl_Bergman}). \citet{SchrempfHobolth17} were able to explicitly derive the approximate neutral multiallelic stationary distribution of allele frequencies as the equilibrium state of a discrete boundary mutation Moran model with a completely general mutation model. Starting from a Wright-Fisher model and passing to the diffusion limit, \citet{BurdenGriffiths18(1)} showed that a Taylor series expansion in the scaled mutation rate can be used to obtain the stationary distribution for mutation models more complex than the parent-independent mutation model. This is remarkable because the equilibrium distribution of such models is not known. Below, we provide an alternative proof. 

Our main results, however, are ML estimators for the scaled mutation parameters of a strand-symmetric mutation model. The two strands of the DNA double helix are held together in anti-parallel orientation by hydrogen bonds. Assuming symmetry between the two strands, the twelve mutation rate parameters between the four bases adenine, thymine, cytosine, and guanine ($A$, $T$, $C$, and $G$) are reduced to six because the transition rates of complementary bases are identical, \eg\  the transition rate from $A$ to $C$ is identical to that from $T$ to $G$. For these six parameters, we provide estimators using expectation-maximization (EM) algorithms that are guaranteed to converge to the global maximum. 

\section{The strand-symmetric mutation-drift model}

At each site, we assume $K=4$ alleles corresponding to the four bases. The evolutionary process is described by a continuous time Markov chain with instantaneous transition rate matrix $\mathbf{Q}$ of $dim(\mathbf{Q})=4$x$4$ and corresponding forward transition probability matrix $\mathbf{P}$. Under the most general mutation model, the off-diagonal entries of $\mathbf{Q}$ are the strictly positive transition rates among the four bases $A$, $T$, $G$, and $C$:
\begin{equation}\label{eq:general_mutation_matrix}
\mathbf{Q} = \begin{tiny}\begin{pmatrix}
 -a_{12}-a_{13}-a_{14}&  a_{12}&  a_{13}& a_{14}\\ 
 a_{21}&  -a_{21}-a_{23}-a_{24}& a_{23} & a_{24}\\ 
 a_{31}&  a_{32}& -a_{31}-a_{32}-a_{34} & a_{34}\\ 
 a_{41}&  a_{42}& a_{43} & -a_{41}-a_{42}-a_{43}\\ 
\end{pmatrix}\end{tiny}\,.    
\end{equation}
Thus the general transition rate matrix has twelve parameters. Finding the unique stationary state for a general transition rate matrix $\mathbf{Q}$ requires scaling the individual entries to ensure they are $\leq 1$. Applying the Perron-Frobenius theorem to the stochastic matrix given by the sum of this scaled transition rate matrix and the identity matrix yields a dominant eigenvalue of zero. All further eigenvalues have a negative real component. Provided the process is irreducible, the eigenvector associated with the zero-valued eigenvalue corresponds to the unique stationary distribution of the process. 

With the strand-symmetric mutation model, similar conclusions can be drawn without invoking the Perron-Frobenius theorem. The number of parameters of $\mathbf{Q}$ is reduced to six because the transition rates between bases are identical on complementary DNA strands. The system is thereby simplified and $\mathbf{Q}$ can be written as in \citet{Lobry95} (note that our $\mathbf{Q}$ is the transpose of Lobry's):
\begin{equation}\label{strand-symmetric Q}
\mathbf{Q} = \begin{pmatrix}
 -a-c-e&  a&  c& e\\ 
 a&  -a-c-e& e & c\\ 
 b&  d& -b-d-f & f\\ 
 d& b &  f& -d-b-f
\end{pmatrix}    
\end{equation}\,.
Again $a,..,f$ are strictly positive and the rows of $\mathbf{Q}$ sum to zero. The leading eigenvalue of $\mathbf{Q}$ is $\lambda_1=0$ with the associated normed eigenvector
\begin{equation}
    v_1=\begin{pmatrix}
    \frac{d+b}{2(b+c+d+e)} & 
    \frac{d+b}{2(b+c+d+e)} & 
    \frac{c+e}{2(b+c+d+e)} & 
    \frac{c+e}{2(b+c+d+e)}
    \end{pmatrix}\,.
\end{equation}
This vector corresponds to the stationary distribution \citep{Lobry95}. Clearly, the probabilities of the four bases at stationarity follow Chargaff's second parity rule: For each of the two DNA strands, the expectations of the proportions of $A$ and $T$ are identical, as are those of $C$ and $G$. Chargaff's second parity rule has been shown to hold for all types of double-stranded DNA except organellar DNA \citep{MitchellBridge06}. 

The second eigenvalue \citep{Lobry95} is  $\lambda_2=-(b+d+c+e)$ with associated eigenvector 
$$
v_2=\begin{pmatrix}
1 & 1 & -1 &-1
\end{pmatrix}\,.
$$ 
This eigenvector contrasts the sum of $A$ and $T$ with the sum of $C$ and $G$. 

The next two eigenvalues ($\lambda_3,\lambda_4$) may have complex components that introduce a probability flow into the system \citep{Lobry95}. However, it must be noted that if all transition rates are similar in magnitude, this potentially imaginary term fluctuates around zero. The stationary distribution given by the leading eigenvector is stable despite the complex eigenvalues because the real components of $\lambda_{2,r;3,r;4,r}$ are strictly negative \citep{Lobry95}. Provided the individual transition rates are indeed similar in magnitude, the real components of $\lambda_{3,r}$ and $\lambda_{4,r}$ are equal to  $-a-f+\frac{\lambda_2}{2}$ and are of the same order of magnitude as $\lambda_2$ (The sign in front of the fraction is incorrect in Lobry's listing of eigenvalues; his further arguments indicate that this is merely a typo). Thus the evolutionary process converges exponentially towards a stationary state at a rate that depends on $\lambda_{2,3,4}$.

On the basis of Lobry's work, we can be confident of a unique stationary distribution given the transition rates among alleles in the strand-symmetric mutation model except in degenerate and thus biologically meaningless cases. 

\section{The stationary distribution of the strand-symmetric mutation-drift model with small scaled mutation rates}

In this section, we derive the stationary distribution for a sample of size $M$ taken from the general multiallelic boundary-mutation Moran model in the limit of small scaled overall mutation rates. It is assumed that the mutation matrix gives rise to a unique stationary distribution. This is a similar approach to that of \citet{BurdenGriffiths18(1)} and recovers the full stationary distribution for the proportions of alleles (their formulas~7-9 in  Theorem~1) \citep[see also][]{SchrempfHobolth17}.

We reparametrize the general mutation matrix as follows:
\begin{equation}
\mathbf{Q}  =\theta\cdot\begin{pmatrix}
 -\alpha_{1}&  \alpha_{12}&  \alpha_{13}& \alpha_{14}\\ 
 \alpha_{21}&  -\alpha_{2}& \alpha_{23} & \alpha_{24}\\ 
 \alpha_{31}&  \alpha_{32}& -\alpha_{3} & \alpha_{34}\\ 
 \alpha_{41}&  \alpha_{42}& \alpha_{43} & -\alpha_{4}\\ 
\end{pmatrix}
\,,    
\end{equation}
where $\alpha_k=\sum_{l=1,l\neq k}^K\alpha_{kl}$. 

Assuming a unique stationary distribution for $\mathbf{Q}$ and recalling the duality between the Moran model and Kingman's coalescent \citep[][chapt.~2]{Etheridge12}, we can use a sampling algorithm proposed by \citet{StephensDonnelly00} to build a genealogical realization of a Moran model of sample size $M$ forwards in time. This sampling algorithm is essentially an urn sampling process. \citet{Hoppe87} first drew the connection between Polya-like urn processes and Kingman's coalescent; \citet{DonellyKurtz96} proposed an algorithm by which one can obtain the realization of a sample from a particle Fleming-Viot process at stationarity forwards in time (The Moran model falls within the attraction of the Fleming-Viot processes). The Stephens-Donnelly algorithm is the continuous version of such particle sampling algorithms:

\begin{enumerate}
    \item Start with a sample of size $m=1$. Randomly select an allele of type $k$ from $1\leq k \leq K$ with probability $\Pr(k\given \vec{ \alpha})$ taken from the stationary distribution characterized by $\mathbf{Q}$. Split immediately into two lineages of this type.
    \item At this point there are $m$ lineages in the ancestry. Wait for an exponentially distributed time with rate $m(m-1+\theta)$ and then select an ancestral lineage at random. Split it with probability $m(m-1+\theta)$, otherwise introduce a mutation.
    \item Hit a predetermined stopping criterion at sample size $M+1$, then go back to the time when there were $M$ ancestral lineages.
\end{enumerate}

\section{Sampling Paths}

\subsection{Taylor Expansion}

Given a sample of alleles drawn with the Stephens-Donnelly algorithm, one can explicitly calculate the probabilities of the possible sampling paths that lead to the observed sample. 

A monomorphic sample of size $M$ is either the result of a sampling path that consists purely of splitting ancestries or of one that has an even number of reversible mutations (which we also take to include the unlikely case of four mutations through all variants back to the original allele). In the former case, the sampling path can be written as follows:
\begin{equation} 
\begin{split}
f_0(\theta) &= \Pr(y_k =M\mid\vec{ \alpha},M,\theta,\text{no mutation})\\
&= \Pr(k\mid\vec{ \alpha})\cdot\frac{1}{1+\alpha_{k}\theta}\cdot\frac{2}{2+\alpha_{k}\theta} \cdots\frac{M-1}{(M-1)+\alpha_{k}\theta}\,.
\end{split}
\end{equation}  
With $\theta \ll 1$ and $\alpha_i$, $\alpha_{ij}$ of order one or smaller, a Taylor expansion of the full sampling path at $\theta=0$ is to first order:
\begin{equation}
\Pr(y_k =M\mid\vec{ \alpha},M,\text{no mutation})=\Pr(k\mid\vec{ \alpha})-\Pr(k\mid\vec{ \alpha})\cdot\sum_{m=2}^{M}\frac{\alpha_{k}\theta}{m-1} + \mathcal{O}(\theta^{2})\,.
\end{equation}

\noindent
Here the full derivation: We have 
\begin{equation*} 
f_0(\theta=0)=\Pr(k\mid\vec{ \alpha})
\end{equation*}
and
\begin{equation*} 
\begin{split}
f_0'(\theta)&=\Pr(k\mid\vec{ \alpha})\cdot\bigg(-\frac{\alpha_{k}}{(1+\alpha_{k}\theta)^{2}} \bigg[ \frac{2}{2+\alpha_{k}\theta}\cdot\frac{3}{3+\alpha_{k}\theta}\cdots\frac{(
M-1)}{(M-1)+\alpha_{k}\theta} \bigg]- \\   
&\qquad -\frac{2\alpha_{k}}{(2+\alpha_{k}\theta)^{2}} \left [ \frac{1}{1+\alpha_{k}\theta}\cdot\frac{3}{3+\alpha_{k}\theta}\cdots\frac{(M-1)}{(M-1)+\alpha_{k}\theta} \right ] - \cdots\\   
&\qquad \cdots-\frac{(M-1)\alpha_{k}}{((M-1)+\alpha_{k}\theta)^{2}} \bigg[ \frac{1}{1+\alpha_{k}\theta}\cdot\frac{2}{2+\alpha_{k}\theta}\cdots\frac{(M-2)}{(M-2)+\alpha_{k}\theta} \bigg ] \bigg)\,,
\end{split}
\end{equation*}
such that
\begin{equation*} 
f_0'(\theta=0)=-\Pr(k\mid\vec{ \alpha})\cdot\sum_{m=2}^{M}\frac{\alpha_{k}}{m-1}\,.
\end{equation*}
Note that $m$ can be interpreted as the sample size at which a mutation could potentially occur.

A sampling path with an even number of mutations that restores the monomorphic condition is unlikely with low mutation rates. Indeed it is easy to show that, in this case, probabilities of such sampling paths are at least of second order in $\theta$ and thus do not contribute to the first order approximation.

Consider now the sampling paths that create polymorphic samples of size $M$: The simplest of these includes one mutation, \eg\  when the sample size $m$ is equal to $y_k+1$: 
\begin{equation}
\begin{split}
   f_1(\theta)&=\Pr(y_k = M - y_l \given \vec{ \alpha},M,\theta, \text{one mutation at $m=y_k+1$})\\
    &=\Pr(k\given \vec{ \alpha})\cdot\frac{1}{1+\alpha_{k}\theta}\cdot\frac{2}{2+\alpha_{k}\theta} \cdots\frac{y_{k}-1}{y_{k}-1+\alpha_{k}\theta}\cdot\frac{\alpha_{kl}\theta}{m-1+\alpha_{k}\theta} \\ &\qquad\times\frac{1}{y_{k}+\frac{y_{k}}{y_{k}+1}\alpha_{k}\theta+ \frac{1}{y_{k}+1}\alpha_{l}\theta}\cdots\frac{y_{l}-1}{M-1+\frac{y_{k}}{M}\alpha_{k}\theta+\frac{y_{l}}{M-y_{k}}\alpha_{l}\theta}
\end{split}
\end{equation}
A Taylor expansion of the probability of this sampling path at $\theta=0$ yields: 
\begin{equation}
\Pr(y_k =M - y_l\given \vec{ \alpha},M, \text{one mutation})= \Pr(k\given \vec{ \alpha})\cdot\frac{\alpha_{kl}}{m-1}\frac{(y_k-1)!(y_m-1)!}{(M-1)!}+ \mathcal{O}(\theta^{2})
\end{equation}

\noindent
The derivation is similar to those above: Set
\begin{equation*}
\begin{split}
    g_y(\theta)&=\Pr(k\given \vec{ \alpha})\cdot\frac{1}{1+\alpha_{k}\theta}\frac{2}{2+\alpha_{k}\theta} \cdots\frac{y_{k}-1}{y_{k}-1+\alpha_{k}\theta}\\
    &\qquad\times
    \frac{1}{y_{k}+\frac{y_{k}}{y_{k}+1}\alpha_{k}\theta+ \frac{1}{y_{k}+1}\alpha_{l}\theta}
    \cdots\frac{y_{l}-1}{M-1+\frac{y_{k}}{M}\alpha_{k}\theta+\frac{y_{l}}{M-y_{k}}\alpha_{l}\theta}\,,
\end{split}
\end{equation*}
such that 
\begin{equation}
\begin{split}
    g_y(\theta=0)&=\Pr(k\given \vec{ \alpha})\frac{(y_k-1)!(y_l-1)!}{(M-1)!}\,.
\end{split}
\end{equation}
Then we have
\begin{equation}
\begin{split}
    f_1(\theta)&=\frac{\alpha_{kl}\theta}{(m-1)+\alpha_{k}\theta}g_y(\theta)\\
    f_1'(\theta)&=\bigg(\frac{\alpha_{kl}\theta}{(m-1)+\alpha_{k}\theta}\bigg)'g_m(\theta)+\frac{\alpha_{kl}\theta}{(m-1)+\alpha_{k}\theta}g_m'(\theta)\,.
\end{split}
\end{equation}
Setting $\theta=0$, we get
\begin{equation}
\begin{split}
    f_1'(\theta=0)&=\frac{\alpha_{kl}}{m-1}g_m(\theta=0)+0\\
    &=\Pr(k\given \vec{ \alpha})\cdot\frac{\alpha_{kl}}{m-1}\frac{(y_k-1)!(y_m-1)!}{(M-1)!}\,.
\end{split}
\end{equation}

The occurrence of two or three mutations that lead to a polymorphic sample with two, three, or four segregating alleles is a theoretical possibility. However, the expansion of the sampling path probabilities would again approximate to zero for the same reason as in the case of (multiple) reversible mutations in a monomorphic sampling path. 

Note that the mutation may in principle occur at any sample size between $2\leq m_k \leq (y_k+1)$. Hence, while the series expansion of the monomorphic sampling paths accounts for the only possible sampling scenario, the expansion of a polymorphic sampling path represents one of several feasible branching structures along the sampling algorithm that result in the given configuration of alleles. We thus need to sum over these possibilities.

\subsection{Sum of Ordered Probabilities}

In the following subsection, we show that the sampling distribution of a polymorphic sampling path, conditional on the mutation occurring at sample size $m$, is beta-binomial. The argument is similar to that in \citet{BurdenGriffiths18(1)}. These authors use a direct result equivalent to the deFinetti density of a Polya urn model. We start from the properties of the sampling path and build up the deFinetti representation.

According to deFinetti's theorem \citep{DeFinetti31}, there exists for every infinite sequence of Bernoulli random variables a probability distribution $F$ on $\left [ 0, 1 \right ]$ so that 
\begin{equation}
\begin{split}
&\Pr(X_1=1,\cdots,X_{k}=1,X_{k+1}=0,\cdots,X_{n}=0)= \\
&\qquad\int_{0}^{1}\nu^k (1-\nu ^{n-k}) dF(\nu) = E\left [ U^k (1-U ^{n-k})\right ]    
\end{split}
\end{equation} 
for a random variable $U$, whereby 
$$
\Pr(U\leq\nu)=F(\nu)
$$ 
and the $(X_{i})_i\leq1$ are i.i.d. when conditioned on a $U$ that fulfills $\Pr(X_i=k)=U$. The following also holds: 
$$
\Pr(S_n=\sum_{i}^{n}X_i=k)=\int_{0}^{1}\binom{n}{k}\nu^k (1-\nu ^{n-k}) dF(\nu)\,.
$$

The joint distribution of the polymorphic sampling path, which is a finite number of draws of alleles, does not change if the order of alleles within the path is changed. We start from an initial allele of type $k$, then there is a mutational event to allele $l$ at a random time point. At this point, $m-1$ alleles of type $k$ are already in the sample. Each subsequent draw can increase the number of sampled alleles of either type by one. This is exactly the specification of the following Polya urn process:
$$
(1,m-1)\begin{pmatrix}
1 & 0\\ 
 0& 1
\end{pmatrix}\,,
$$
which in our case runs until we have $y_l-1$ alleles of type $l$ and $y_k-m=M-y_l-m$ alleles of type $k$. By noting that the probabilities of specific sampling paths are sequences of beta distributed moments and recalling Hausdorff's moment problem \citep{Hausdorff21(1),Hausdorff21(2)}, it follows that $$U\sim B(1,m-1)\,.$$

Therefore, the probability of a process that yields $y$ alleles of type $l$ in a sample size of $M-m$ (denoted by $S_{M-m}$) is given by the following:
\begin{equation}
    \begin{split}
        &\Pr(S_{M-m}\given M,m,1)\\
        &=\int_{0}^{1}\binom{M-m}{y_l-1}\nu^{y_l-1}(1-\nu)^{M-y_l-m}(1-\nu)^{m-1}\nu^0\frac{\Gamma(m)}{\Gamma(m-1)\Gamma(1)} d\nu\\
        &=\binom{M-m}{y_l-1}\frac{\Gamma(M-y_l-m+m-1)\Gamma(y_l)}{\Gamma(M-y_l-m+m-1+y_l)}\frac{\Gamma(m)}{\Gamma(m-1)\Gamma(1)}\\
        &=\binom{M-m}{y_l-1}\frac{\Gamma(M-y_l)\Gamma(y_l)}{\Gamma(M)}\frac{\Gamma(m)}{\Gamma(m-1)\Gamma(1)}\\
        &=\binom{M-m}{y_l-1}\frac{\Gamma(M-y_l)\Gamma(y_l)}{\Gamma(M)}\frac{(m-1)!}{(m-2)!(1-1)!}\\
        &=\binom{M-m}{y_l-1}(m-1)\frac{\Gamma(M-y_l)\Gamma(y_l)}{\Gamma(M)}\,.
    \end{split}
\end{equation}

For a given proportion $y_l$ of mutant alleles, we need to sum over all possible mutation points, \ie\  $2\leq m \leq (M-y_l+1)$. Given that the first allele sampled at sample size $m=1$ is of type $k$, we have:
\begin{equation}
    \begin{split}
        \Pr(y_l\given M,\vec{\alpha},k,\theta\to 0)&=\sum_{m=2}^{M-y_l+1}\binom{M-m}{y_l-1}(m-1)\frac{\Gamma(M-y_l)\Gamma(y_l)}{\Gamma(M)} \frac{\alpha_{kl}\theta}{m-1}+\mathcal{O}(\theta^{2})\\
        &=\alpha_{kl}\theta\sum_{m=2}^{M-y_l+1}\binom{M-m}{y_l-1}\frac{\Gamma(M-y_l)\Gamma(y_l)}{\Gamma(M)}+\mathcal{O}(\theta^{2})\\
        &=\frac{\alpha_{kl}\theta}{y_l}+\mathcal{O}(\theta^{2})\,.
    \end{split}
\end{equation}
Line three follows from line two by repeatedly applying the identity:
\begin{equation}
\begin{split}
    \binom{N}{x}&=\binom{N-1}{x-1}+\binom{N-1}{x}\Leftrightarrow \binom{N-1}{x-1}=\binom{N}{x}-\binom{N-1}{x}\,,
\end{split}
\end{equation}
from which it follows that
\begin{equation}\label{eq:binom_sum}
\begin{split}
   &\sum_{m=2}^{M-y_l+1}\binom{M-m}{y_l-1}\\
   &=\binom{M-2}{y_l-1}+\binom{M-3}{y_l-1}+\dots+\binom{y_1}{y_l-1}+\binom{y_l-1}{y_l-1}\\
   &=\binom{M-1}{y_l}-\binom{M-2}{y_l}+\binom{M-2}{y_l}-\binom{M-3}{y_l}+\\
   &\qquad\dots+\binom{y_l+1}{y_l}-\binom{y_l}{y_l}+\binom{y_l}{y_l}-\binom{y_l-1}{y_l}\\
   &=\binom{M-1}{y_l}\,,
\end{split}
\end{equation}
since $\binom{y_l-1}{y_l}=0$. Note that equation~(\ref{eq:binom_sum}) also appears in \citet{BurdenGriffiths18(1)}, but is solved differently. 

This provides an alternative proof of Theorem~1 in \citet{BurdenGriffiths18(1)} and concludes the characterization of the polymorphic sampling paths.

\section{Stationary Distribution}

On slightly reparametrizing the results of the previous two subsections, the probability of obtaining the overall configuration $y$ of alleles can be concisely formulated: Given a fixed sample size $M < e ^{\frac{1}{\alpha_{k}\theta}}$, mutation biases of maximal order $1$, and a small scaled overall mutation rate $\theta= N\mu<0.1$, the stationary allelic configuration $Y$ to the first order of $\theta$ is the distribution $\Pr(y\mid M,\vec{\alpha},\theta )$: 
\begin{equation}\label{eq:general_stationary_distribution}
  \begin{cases}
  \Pr(k\mid\vec{ \alpha})-\Pr(k\mid\vec{ \alpha})\cdot\sum_{k\neq l}^K\alpha_{k}\theta H_{M-1} + \mathcal{O}(\theta^{2}) & y_k=M,y_{l\neq k}=0 \\
  \Pr(k\mid\vec{ \alpha})\frac{\alpha_{kl}\theta}{y_l}+ \Pr(l\mid\vec{ \alpha})\frac{\alpha_{lk}\theta}{y_k} + \mathcal{O}(\theta^{2}) & 1\leq y_k,y_{l\neq k}\leq M-1\\
  &\text{and } y_l+y_k=M\,.
\end{cases}
\end{equation}
All other possibilities have probabilities of at most $\mathcal{O}(\theta^{2})$ and thus do not contribute to the distribution.
This stationary distribution is equivalent to that derived by \citet{BurdenGriffiths18(1)}, as expected.

\section{Maximum likelihood (ML) estimators of the scaled mutation rate parameters}

Starting from the general stationary distribution and recalling that we are certain of a unique stationary distribution of $\mathbf{Q}$ in the case of strand symmetry (\ie\  we can build a sample of size $M$ using the Stephens-Donnelly algorithm if we assume strand symmetry), we now aim to infer the six parameters of the strand-symmetric boundary-mutation Moran model given site frequency spectrum data. 

\begin{figure}[h]
\centering
\includegraphics[scale=0.4]{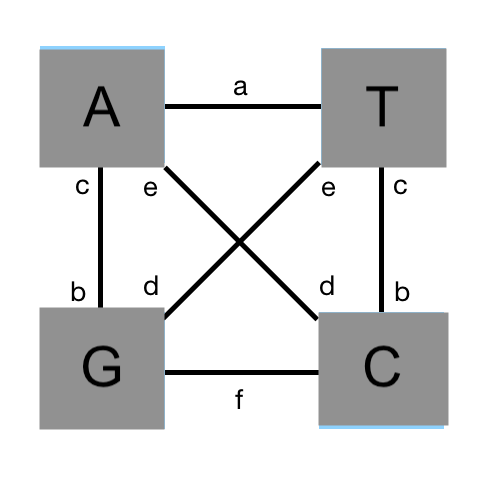}
\caption{Transition Rates for the Strand Symmetric Model}
\end{figure}

Recalling the parametrization for $\mathbf{Q}$ in formula~(\ref{strand-symmetric Q}) as visualized in Fig~1, we set $\theta=-\lambda_2=b+d+c+e$. Then $\alpha_1=\frac{a+c+e}{b+d+c+e}$, $\alpha_{12}=\frac{a}{b+d+c+e}$, $\alpha_{13}=\frac{c}{b+d+c+e}$, $\alpha_{14}=\frac{e}{b+d+c+e}$, \etc. 

The full stationary distribution for the strand-symmetric model is then: 
\begin{equation}\label{eq:stationary_distribution_strand_symmetric}
\begin{split}
&\Pr(y\mid M,\vec{\alpha},\theta)=\\
&\begin{cases}
  \frac{b+d}{2(b+d+c+e)}-\frac{b+d}{2(b+d+c+e)}(a+c+e) H_{M-1} + \mathcal{O}(\theta^{2}) & y_A=M,y_{T,G,C}=0 \\
  \frac{b+d}{2(b+d+c+e)}-\frac{b+d}{2(b+d+c+e)}(a+c+e) H_{M-1} + \mathcal{O}(\theta^{2}) & y_T=M,y_{A,G,C}=0 \\
  \frac{c+e}{2(b+d+c+e)}-\frac{c+e}{2(b+d+c+e)}(b+d+f) H_{M-1} + \mathcal{O}(\theta^{2}) & y_G=M,y_{A,T,C}=0 \\
   \frac{c+e}{2(b+d+c+e)}-\frac{c+e}{2(b+d+c+e)}(b+d+f) H_{M-1} + \mathcal{O}(\theta^{2}) & y_C=M,y_{A,T,G}=0 \\
   \frac{b+d}{2(b+d+c+e)}\frac{a}{y_A}+\frac{b+d}{2(b+d+c+e)}\frac{a}{y_T}+ \mathcal{O}(\theta^{2})  & 1\leq y_A,y_T\leq M-1, y_A+y_T=M \\
   \frac{b+d}{2(b+d+c+e)}\frac{e}{y_A}+\frac{c+e}{2(b+d+c+e)}\frac{d}{y_C}+ \mathcal{O}(\theta^{2})  & 1\leq y_A,y_C\leq M-1, y_A+y_C=M \\
   \frac{b+d}{2(b+d+c+e)}\frac{c}{y_A}+\frac{c+e}{2(b+d+c+e)}\frac{b}{y_G}+ \mathcal{O}(\theta^{2})  & 1\leq y_A,y_G\leq M-1, y_A+y_G=M \\
   \frac{b+d}{2(b+d+c+e)}\frac{c}{y_T}+\frac{c+e}{2(b+d+c+e)}\frac{b}{y_C}+ \mathcal{O}(\theta^{2})  & 1\leq y_T,y_C\leq M-1, y_T+y_C=M \\
   \frac{b+d}{2(b+d+c+e)}\frac{e}{y_T}+\frac{c+e}{2(b+d+c+e)}\frac{d}{y_G}+ \mathcal{O}(\theta^{2})  & 1\leq y_T,y_G\leq M-1, y_T+y_G=M \\
   \frac{c+e}{2(b+d+c+e)}\frac{f}{y_G}+\frac{c+e}{2(b+d+c+e)}\frac{f}{y_C}+ \mathcal{O}(\theta^{2})  & 1\leq y_G,y_C\leq M-1, y_G+y_C=M \,.
\end{cases}
\end{split}
\end{equation}

In the next subsections, we will derive ML estimators for the parameter vector $(a,b,c,d,e,f)$ in the distribution~(\ref{eq:stationary_distribution_strand_symmetric}) by linear transformation.

\subsection{Biallelic Model}

Grouping the bases $A$ and $T$ as well as $G$ and $C$ together results in a biallelic model  \citep{Vogl14}: If one interprets the allelic state $k$ as representing the grouped state $A-T$ including those sites polymorphic for $A$/$T$ and the allelic state $l$ as representing the $G-C$ alleles and the polymorphic $G/C$ sites, $\alpha_{kl}$ and $\alpha_{lk}$ become equal to $\alpha_{k}$ and $\alpha_{l}$ respectively (see Figure 2). 

\begin{figure}[ht]
\centering
\includegraphics[scale=0.4]{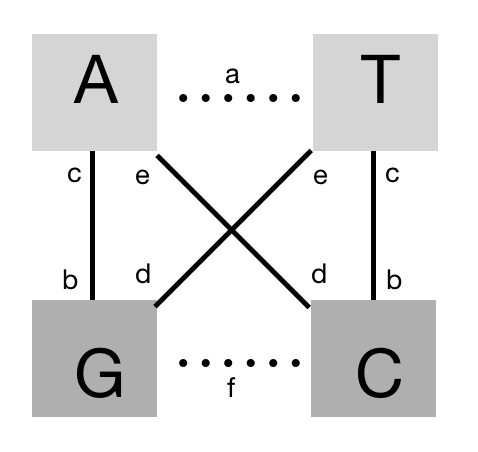}
\caption{Transition Rates for the Biallelic Strand Symmetric Model}
\end{figure}

In particular, this biallelic model enables us to infer $\Pr(k =AT\mid\vec{ \alpha})=\frac{b+d}{b+d+c+e}$ and $\Pr(l =GC\mid\vec{ \alpha})=\frac{e+c}{b+d+c+e}$. Using the frequencies of $A$ and $T$, one can then infer $a$ using a symmetric version of the biallelic model and similarly infer $f$ from the frequencies of $G$ and $C$. The remaining transition rates can be estimated via the expectation-maximization algorithm.

Following \citet{Vogl14}, we will first find the ML estimators for $\Pr(k =AT\mid\vec{ \alpha})=\frac{b+d}{b+d+c+e}=\beta$ and $\Pr(l =GC\mid\vec{ \alpha})=\frac{e+c}{b+d+c+e}=1-\beta$ using $\alpha_{13}+\alpha_{14}=1-\beta$ and starting from the following biallelic version of \ref{eq:general_stationary_distribution}:
\begin{equation}
\Pr(y\mid M,\vec{\alpha},\theta)=
\begin{cases}
  \beta - \beta(1-\beta)\theta H_{M-1} + \mathcal{O}(\theta^{2}) & y_k=M,y_{l\neq k}=0 \\
  \beta(1-\beta)\theta\bigg(\frac{1}{y_l} + \frac{1}{y_k}\bigg) + \mathcal{O}(\theta^{2}) & 1\leq y_k,y_{l\neq k}\leq M-1,\\
  &y_l+y_k=M\\
   (1-\beta) - \beta(1-\beta)\theta H_{M-1} + \mathcal{O}(\theta^{2}) & y_k=0,y_{l\neq k}=M\,.
\end{cases}
\end{equation}
The estimators are to be expressed via site frequency data: We assume knowledge of the allele frequency counts at $L < \infty $ loci indexed as $1\leq l \leq L$ and further assume that these counts are based on a sample of size $M$ at each locus (the constant sample size merely simplifies the notation and calculations and is not a theoretical necessity). $L_0$ is the number of loci at which there are no occurrences of the state $A-T$ (\ie\  the number of sites monomorphic for $G$ or $C$, or $G/C$ polymorphic), and  $L_M$ is the number of loci either monomorphic for $A$ or $T$, or $A/T$ polymorphic. The remaining $L_y$ with $L_p=\sum_{y=1}^{M-1}L_y$ are the observed polymorphic counts between the $A-T$ and $G-C$ states. We define 
$$
\gamma= \beta(1-\beta)\theta=\frac{(b+d)(e+c)}{b+d+c+e}
$$ 
to efficiently handle products of random variables.

The likelihood of the site frequency spectrum then becomes:
 \begin{equation}
 \begin{split}
&\Pr(L_0\dots L_M \mid M,\vec{\alpha},\gamma)=\\
&\qquad\frac{L!}{\prod_{y=0}^{M}L_y!}(\beta -\gamma H_{M-1})^{L_M}
(2\beta \gamma H_{M-1})^{L_p}
((1-\beta)-\gamma H_{M-1})^{L_0}\,,
 \end{split}
\end{equation}
whereby $H_m$ again denotes the $m$th harmonic number. 

Then the log-likelihood can be written as proportional to:
\begin{equation}
\begin{split}\label{eq:biallelic log-lh}
\log(\Pr(L_0\dots L_M \mid M,\vec{\alpha},\gamma))&= const+  
 {L_M}\log(\beta -\gamma H_{M-1})\\
&\qquad+{L_p}\log(2\beta \gamma H_{M-1})+ {L_0}\log((1-\beta)-\gamma H_{M-1}) \,.
\end{split}
\end{equation}
With  asymmetric mutation rates between the two classes, we can take the derivative of the log-likelihood by $\beta$:
\begin{equation}
\begin{split}
\frac{-L_0}{(1-\beta)-\gamma H_{M-1}}+\frac{L_M}{\beta-\gamma H_{M-1}} & \overset{!}{=} 0\\
-L_0 \beta + L_0\gamma H_{M-1} + L_M - L_M \beta - L_M\gamma H_{M-1} &= 0 \,.
\end{split}
\end{equation}
It follows that 
\begin{equation}
 \beta = \frac{L_0\gamma H_{M-1}+L_M(1-\gamma H_{M-1})}{L_0+L_M}\,,
\end{equation}
and therefore
\begin{equation}
\begin{split}
  \beta - \gamma H_{M-1} &= \frac{-L_0\gamma H_{M-1}-L_M\gamma H_{M-1}+L_0\gamma H_{M-1} +L_M(1-\gamma H_{M-1})}{L_0+L_M}\\
  &= \frac{L_M(1-2\gamma H_{M-1})}{L_0+L_M}\,,
\end{split}
\end{equation}
and similarly
\begin{equation}
 (1-\beta) - \gamma H_{M-1} = \frac{L_0(1-2\gamma H_{M-1})}{L_0+L_M}\,.
\end{equation}
Taking the derivative of the log-likelihood by $2\gamma H_{M-1}$ and substituting we get:
\begin{equation}
\begin{split}
&\log(\Pr(L_0\dots L_M \mid M,\vec{\alpha},\gamma)=\frac{-L_M(L_0+L_M)\frac{L_M}{L_0+L_M}}{L_M(1-2\gamma H_{M-1})}+\frac{L_p}{2\gamma H_{M-1}}\\
&\qquad-\frac{L_0(L_0+L_M)\frac{L_0}{L_0+L_M}}{L_0(1-2\gamma H_{M-1})} 
 \overset{!}{=}0\\
&-(L_0+L_M)2\gamma H_{M-1} + L_p(1-2\gamma H_{M-1}) = 0 \,.
\end{split}
\end{equation}
It follows that 
\begin{equation}
\widehat{\gamma }= \frac{L_p}{2(L_0+L_M+ L_p)H_{M-1}}\,.
\end{equation}
This corresponds to the estimator in formula~(36) in  \citep{Vogl14}. Analogously, we get:
\begin{equation}
\beta -\widehat{\gamma} H_{M-1}= \frac{L_M}{(L_0+L_M+ L_p)}\,.
\end{equation}

Similarly, we have
\begin{equation}
(1-\beta)-\widehat{\gamma} H_{M-1}= \frac{L_0}{(L_0+L_M+ L_p)}
\end{equation}
respectively yielding 
\begin{equation}
\widehat{\Pr(k =AT\mid\vec{ \alpha})}=\widehat{\beta}= \frac{L_M +\frac{L_p}{2} }{(L_0+L_M+ L_p)}\,,
\end{equation}
which corresponds to the estimator in formula~(37) in  \citep{Vogl14}, and
\begin{equation}
\widehat{\Pr(k =GC\mid\vec{\alpha})}= 1-\widehat{\beta}=\frac{L_0 +\frac{L_p}{2} }{(L_0+L_M+ L_p)}\,.
\end{equation}

Thus we provide ML estimators of the parameters of the combined states $A-T$ and $G-C$ in the biallelic model resulting from combining $A-T$ and $C-G$ under stationarity.

\subsection{Symmetric Biallelic Model}
    
\begin{figure}[h]
\centering
\includegraphics[scale=0.4]{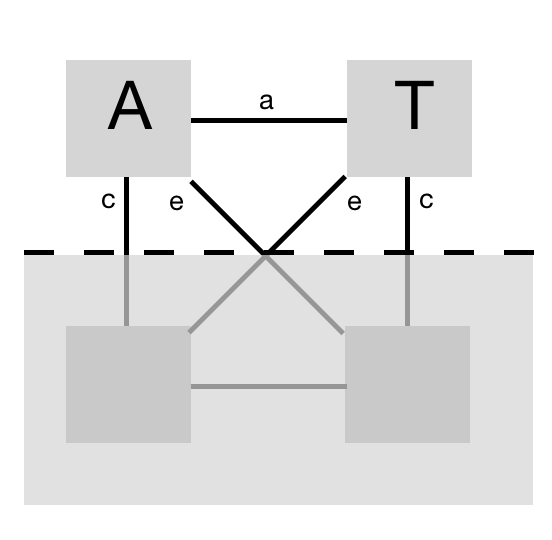}
\caption{Transition Rates for the Symmetric Biallelic Strand Symmetric Model}
\end{figure}

Next we recover estimators for the transition rates $a$ and $f$ from symmetric biallelic systems \citep[see][]{BurdenTang16}. Let us focus on $a$ (see Figure 3): Recalling Chargaff's second parity rule and its correspondence with the stationary distribution of base frequencies, we can write  
$$
\Pr(k=A\mid\vec{ \alpha})= \Pr(k=T\mid\vec{ \alpha})=\frac{\Pr(k =AT\mid\vec{ \alpha})}{2}=\frac{\beta}{2}=\frac{b+d}{2(b+d+c+e)}.
$$ 
Furthermore, we know that
$$
\gamma = \beta(1-\beta)\theta = \frac{b+d}{b+d+c+e}(c+e)\,,
$$ 
from which it follows that
$$
\frac{\beta}{2}(a+c+e) = \frac{\beta}{2}a+\frac{\gamma}{2}
$$
Subsequently the distribution~(\ref{eq:general_stationary_distribution}) can be rewritten for a symmetric biallelic A-T model using the previously estimated parameters:
\begin{equation}
\Pr(y\mid M,a,\beta,\gamma)=
\begin{cases}
  \frac{\beta}{2}- \frac{\beta}{2} a H_{M-1}-\frac{\gamma}{2}H_{M-1} + \mathcal{O}(\theta^{2}) & y_A=M,y_{T}=0\\ &\text{ or }y_A=0,y_{T}=M\\
  \beta a \bigg(\frac{1}{y_A}+\frac{1}{y_T}\bigg) + \mathcal{O}(\theta^{2}) & 1\leq y_A,y_{T}\leq M-1,\\ &y_A+y_T=M\end{cases}
\end{equation}
This yields the following log-likelihood of the site frequency spectrum: 
\begin{equation}
\begin{split}\label{eq:symmetric biallelic log-lh}
&\log(\Pr(L_0,\dots,L_M \mid M,a,\beta,\gamma))= const+  
 {L_M}\log\bigg(\frac{\beta}{2}- \frac{\beta}{2} a H_{M-1} - \frac{\gamma}{2}H_{M-1}\bigg)\\
&\qquad+{L_p}\log(\beta a H_{M-1})+ {L_0}\log\bigg(\frac{\beta}{2}- \frac{\beta}{2} a H_{M-1} - \frac{\gamma}{2}H_{M-1}\bigg) \\ 
\end{split}
\end{equation}
Taking the derivative by $a$, one can calculate the ML estimator for $a$ as follows:
\begin{equation}
\begin{split}
\frac{-L_M \frac{\beta}{2} H_{M-1}}{\frac{\beta}{2}- \frac{\beta}{2} a H_{M-1} - \frac{\gamma}{2}H_{M-1}}
+\frac{L_p \beta H_{M-1}}{\beta a H_{M-1}} - 
\frac{L_0 \frac{\beta}{2} H_{M-1}}{\frac{\beta}{2}- \frac{\beta}{2} a H_{M-1} - \frac{\gamma}{2}H_{M-1}} &\overset{!}{=} 0\\
\frac{-L_M }{\frac{1}{H_{M-1}} - a - (1-\beta)\theta}
+\frac{L_p}{ a } - 
\frac{L_0 }{\frac{1}{H_{M-1}} - a - (1-\beta)\theta} &= 0\\
 \frac{L_p\bigg(\frac{1}{H_{M-1}} - (1-\beta)\theta\bigg)}{(L_M + L_p + L_0)} &= a\\
\end{split}
\end{equation}
From earlier we have
$$
(1-\beta)\theta = \frac{\gamma}{\beta}\,,
$$
such that 
$$
\widehat{(1-\beta) \theta} = \frac{\hat{\gamma}}{\hat{\beta}} = \frac{L_p}{2(L_M+\frac{L_p}{2})H_{M-1}}\,,
$$
and furthermore
$$
\frac{1}{H_{M-1}} - \widehat{(1-\beta) \theta} = \frac{2L_M}{2H_{M-1}(L_M + \frac{L_p}{2})}\,.
$$
As a result the ML estimator for $a$ becomes:
\begin{equation}
\begin{split}
 \hat{a}& =
\frac{L_pL_M}{(L_M + L_p + L_0)(L_M + \frac{L_p}{2})H_{M-1}} \\
\end{split}
\end{equation}
Analogously, we get
\begin{equation}
\begin{split}
 \hat{f}& =
\frac{L_pL_0}{(L_M + L_p + L_0)(L_0 + \frac{L_p}{2})H_{M-1}} \,.
\end{split}
\end{equation}

\subsection{Expectation-Maximization Algorithm}

The remaining transition rates $(b,d,e,c)$ cannot be disentangled and expressed individually through reparametrization in the same way as $a$ and $f$. Instead, we will use the expectation-maximization (EM) algorithm \citep{DempsterLairdRubin77} to obtain estimators of these transition rates as sequential updates that cycle through the parameter pairs $d,e$ and $b,c$, respectively.

\begin{figure}[h]
\centering
\includegraphics[scale=0.4]{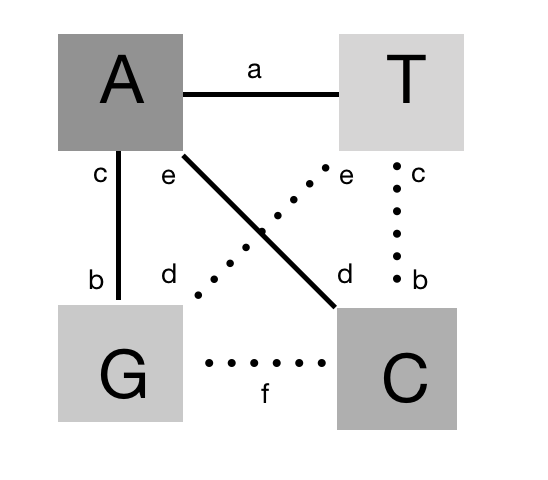}
\caption{Transition Rates for the Diagonally Symmetric Biallelic Strand Symmetric Model}
\end{figure}

Let us focus on a biallelic A-C system (see Figure 4) and rewrite \ref{eq:general_stationary_distribution} to obtain the appropriate stationary distribution:
\begin{equation}
\Pr(y\mid M,e,d,\beta,\theta)=
\begin{cases}
  \frac{\beta}{2} - \frac{\beta}{2} (a+c+e) H_{M-1} + \mathcal{O}(\theta^{2}) & y_A=M,y_C=0 \\
   \frac{\beta}{2}\frac{e}{y_A} + \bigg(\frac{1-\beta}{2}\bigg)\frac{d}{y_C}+ \mathcal{O}(\theta^{2}) & 1\leq y_A,y_C\leq M-1,\\ & y_A+y_C=M\\
   \bigg(\frac{1-\beta}{2}\bigg) - \bigg(\frac{1-\beta}{2}\bigg)(d+b+f)  H_{M-1}  + \mathcal{O}(\theta^{2}) & y_C=M,y_A=0\,.
\end{cases}
\end{equation}

Let $L_i$ denote the number of polymorphic samples in the frequency spectrum, whereby $y_A=i$ and $y_C=M-i$. In order to distinguish between mutations of different directions, we introduce the auxiliary variable $z_i$ ($0\leq z_i\leq L_i$) that counts the mutations from $A$ to $C$. Conversely, $(L_i-z_i)$ counts the mutations from $C$ to $A$. This yields the complete data log-likelihood:
\begin{equation}
\begin{split}
&\Pr(y,z_i\given L_0,\dots,L_i,\dots,L_{M}, \dots)= L_M \log\bigg(\frac{\beta}{2} - \frac{\beta}{2} (a+c+e) H_{M-1}\bigg)\\
&\qquad +\sum_{i=1} ^{M-1} z_i \log\bigg(\frac{\beta}{2}\frac{e}{i}\bigg) +\sum_{i=1} ^{M-1}(L_i-z_i)\log\bigg(\frac{1-\beta}{2}\frac{d}{M-i}\bigg)\\
&\qquad +L_0 \log\bigg(\frac{1-\beta}{2} - \frac{1-\beta}{2}(d+b+f)  H_{M-1}\bigg)\,.
\end{split}
\end{equation}
The expectation of $z_i$ corresponds to the mean of a binomial distribution with sample size $L_i$ and $p_i=\frac{\frac{e}{i}}{\frac{e}{i}+\frac{d}{M-i}}$:
\begin{equation}
    \E(z_i\given L_i,e^t,d^t)=L_i\frac{\frac{e^t}{i}}{\frac{e^t}{i}+\frac{d^t}{M-i}}\,.
\end{equation}

The expectation step of the EM-algorithm constitutes taking the expectation of the full log-likelihood and noting that only the part $\operatorname{Q}(d,e\given d^t,e^t)=\E(\Pr(y,z_i\given L_0,\dots,L_i,\dots,L_{M}, \dots))$ of the function needs to be maximized:
\begin{equation}
\begin{split}
&\operatorname{Q}(d,e\given d^t,e^t)=const + L_M \log\bigg(1 - (a+c+e) H_{M-1}\bigg)\\
&\qquad +\sum_{i=1} ^{M-1} L_i \frac{\frac{e^t}{i}}{\frac{e^t}{i}\frac{d^t}{M-i}} \log(e) +\sum_{i=1} ^{M-1} L_{M-i} \frac{\frac{d^t}{M-i}}{\frac{e^t}{i}
+\frac{d^t}{M-i}}\log(d)\\
&\qquad +L_0 \log\bigg(1 - (d+b+f)  H_{M-1}\bigg)\,.
\end{split}
\end{equation}

To calculate the parameter updates according to the maximization step, we must take the appropriate derivatives of $\operatorname{Q}(d,e\given d^t,e^t)$. For the calculation of $e^{t+1}$, we take the derivative by $e$ and set it to zero:
\begin{equation}
\begin{split}
\frac{d}{de}\operatorname{Q}(d,e\given d^t,e^t)&\overset{!}{=} 0\\
0&=\frac{-L_M H_{M-1}}{1- (a+c+e)H_{M-1}}
+\sum_{i=1}^{M-1}\frac{\frac{L_i\frac{e^t}{i}}{\frac{e^t}{i}+\frac{d^t}{M-i}}}{e}\\
0&= \frac{-L_M H_{M-1}}{1-(a+(1-\beta)\theta)H_{M-1}}
+\frac{\sum_{i=1}^{M-1}L_i\frac{\frac{e^t}{i}}{\frac{e^t}{i}+\frac{d^t}{M-i}}}{e}\\
e&=\frac{1-(a+(1-\beta)\theta)H_{M-1}}{L_M H_{M-1}}\sum_{i=1}^{M-1}L_i\frac{\frac{e^t}{i}}{\frac{e^t}{i}+\frac{d^t}{M-i}}\,,
\end{split}
\end{equation}
where we substituted $(1-\beta)\theta$ for $c+e$ to obtain the third line.

Substituting the ML estimators for $a$ and $(1-\beta)\theta$ in the numerator of the first factor yields:
\begin{equation}
\begin{split}
(\hat{a}+\widehat{(1-\beta)\theta})H_{M-1}&= \frac{L_pL_M}{(L_M + L_p + L_0)(L_M + \frac{L_p}{2})}+\frac{L_p}{2(L_M + \frac{L_p}{2})}\\
&= \frac{3L_pL_M + L_pL_p + L_pL_0}{2(L_M + L_p + L_0)(L_M + \frac{L_p}{2})}\,.
\end{split}
\end{equation}
Furthermore, we have:
\begin{equation}
\begin{split}
1-(\hat{a}+\widehat{(1-\beta)\theta})H_{M-1}
&=\frac{2L_0L_M + 2L_ML_M}{2(L_M + L_p + L_0)(L_M + \frac{L_p}{2})}\\
&= \frac{L_M(L_0+L_M)}{(L_M + L_p + L_0)(L_M + \frac{L_p}{2})}\,.
\end{split}
\end{equation}
Therefore, we get:
\begin{equation}
\begin{split}
\hat{e}= \frac{(L_0+L_M)}{(L_M + L_p + L_0)(L_M + \frac{L_p}{2}) H_{M-1}}\sum_{i=1}^{M-1}L_i\frac{\frac{e^t}{i}}{\frac{e^t}{i}+\frac{d^t}{M-i}}\,.
\end{split}
\end{equation}

Similarly, one obtains $d^{t+1}$ by taking the derivative of $Q(d,e\given d^t,e^t)$ by $d$ and setting it to zero.

The overall iteration scheme is then given by:
\begin{equation}
    \begin{split}
    e^{t+1} &=\frac{(L_0+L_M)}{(L_M + L_p + L_0)(L_M + \frac{L_p}{2}) H_{M-1}}\sum_{i=1}^{M-1}L_i\frac{\frac{e^t}{i}}{\frac{e^t}{i}+\frac{d^t}{M-i}}\\
    d^{t+1} &=\frac{(L_0+L_M)}{(L_M + L_p + L_0)(L_0 + \frac{L_p}{2}) H_{M-1}}\sum_{i=1}^{M-1}L_i\frac{\frac{d^t}{i}}{\frac{e^t}{i}+\frac{d^t}{M-i}}\,.
    \end{split}
\end{equation}

Considering a biallelic A-G system and using the EM-algorithm analogously, the parameter updates for $c$ and $b$ can also be determined:    
\begin{equation}
    \begin{split}
    c^{t+1} &=\frac{(L_0+L_M)}{(L_M + L_p + L_0)(L_M + \frac{L_p}{2}) H_{M-1}}\sum_{i=1}^{M-1}L_i\frac{\frac{c^t}{i}}{\frac{c^t}{i}+\frac{b^t}{M-i}}\\
    b^{t+1} &=\frac{(L_0+L_M)}{(L_M + L_p + L_0)(L_0 + \frac{L_p}{2}) H_{M-1}}\sum_{i=1}^{M-1}L_i\frac{\frac{b^t}{i}}{\frac{c^t}{i}+\frac{b^t}{M-i}}\,.
    \end{split}
\end{equation}

For both pairs of parameters, cyclical calculation of estimators guarantees convergence towards a local maximum of the marginal likelihood by properties of the EM-algorithm. Furthermore: $\Pr(y\given L_0,\dots,L_i,\dots,L_{M}, \dots)$ is the distribution of counts of each type of segregating  allele. The configurations of the types of alleles themselves are constructed via Polya-like urn processes. As such, the marginal likelihood $\Pr(y\given L_0,\dots,L_i,\dots,L_{M}, \dots)$ takes the form of a Dirichlet-multinomial distribution, which is known to be unimodal \citep{LevinReeds77}. Therefore the estimators determined via the expectation-maximization algorithm converge towards the global optimum of the marginal distribution. 

\section{Summary}

An alternative derivation of the distribution of a sample of size $M$ is derived given a general mutation model under the assumptions of small scaled mutation rates and mutational equilibrium. Further assuming a standard four allele DNA model with strand-symmetric mutation on complementary DNA strands and available site frequency spectrum data, ML estimators for the six scaled mutation parameters are determined. This is the first time such estimators are provided for a mutation model more complex than the parent-independent mutation model. 

\section*{Acknowledgments}
This article is a condensed version of LCM's master's thesis supervised by CV. CV wants to thank Juraj Bergman and Sandra Peer for discussions. CV's research is supported by the Austrian Science Fund (FWF): DK W1225-B20; LCM's by the School of Biology at the University of St.Andrews. 

\section*{References}

\bibliography{refs}  

\end{document}